\documentstyle[12pt]{article}

\begin{document}

\vspace{2mm}

\begin{flushright}
Preprint MRI-PHY/97-2 \\
hep-th/9701154
\end{flushright}

\vspace{2ex}

\begin{center}
{\large \bf Can String Theory Avoid Cosmological 
Singularities?}

\vspace{6mm}
{\large  S. Kalyana Rama}
\vspace{3mm}

Mehta Research Institute, 10 Kasturba Gandhi Marg, 

Allahabad 211 002, India. 

\vspace{1ex}
email: krama@mri.ernet.in \\ 
\end{center}

\vspace{4mm}

\begin{quote}
ABSTRACT. 
We consider the effective action for strings and describe 
in detail the evolution of a four dimensional homogeneous 
isotropic universe with matter included. We find that 
the evolution, which is singular in general, becomes 
singularity free if during certain Phase of the evolution, 
when the scale factor increases and the effective string 
coupling becomes strong, the universe is dominated by 
solitonic $p$-branes, $p = 0$ and/or $- 1$, or by `matter' 
for which $(pressure) \le - \frac{1}{\sqrt{3}} \; 
(density)$. The mechanism in the case of branes is 
reminiscent of the recently discovered field theory 
mechanism where heavy states become light and resolve 
the moduli space singularities. 
\end{quote}

\newpage

\vspace{2ex}

{\bf 1.} 
In cosmology the evolution of the universe generically 
has a singularity in the past, usually of big bang 
type as in standard cosmology. The energy scale 
involved is Planckian and quantum gravity effects 
are expected to play an important role in resolving 
the singularities. String theory is a leading candidate 
for a theory of quantum gravity and, more generally, of 
Physics at Planck scale. Therefore, it is natural to look 
for a resolution of the cosmological singularity within 
the context of string theory. 

There have been numerous attempts to resolve 
the singularity within string theory. In 
\cite{bvafa}-\cite{tseyt} stringy T-duality symmetry 
is used to relate small size to large size where winding 
modes prevent the expansion of the universe to $\infty$ 
and thus, by T-duality, its collapse to zero size. 
Veneziano and collaborators have used this symmetry and 
developed the `pre big bang cosmology' which contains 
a superinflationary branch and a standard cosmology 
branch \cite{ven1,ven2}. A `graceful exit' from 
the first to the second branch will then result in 
a singularity free evolution of the universe. So far, 
however, the graceful exit has been problematic 
\cite{ven2} (see \cite{jmu} for a recent attempt to 
solve this problem). 

Cosmological solutions to one-loop corrected string 
effective action have been studied in \cite{art,em} 
and it is shown \cite{em} that non singular solutions 
exist when the spatial curvature is positive. Non 
singular solutions are also found upon including higher 
derivative terms that incorporate `limiting curvature 
hypothesis' \cite{brand}. 

Cosmological solutions have been studied in M-theory 
\cite{m} with non trivial Ramond-Ramond sector fields 
present. Decomposing the space into a set of maximally 
symmetric subspaces, a class of singularity free 
solutions have been obtained when the spatial curvature 
is positive \cite{burt}. These solutions are related to 
black $p$-brane solutions in a regime where space and 
time reverse their roles \cite{burt,behrndt}. Using such 
a role reversal of space and time, a scenario has been 
proposed recently \cite{wilczek} where big 
bang/crunch singularities are resolved at non singular 
horizons of higher dimensional quasi black hole solutions 
or, plausibly, at Dirichlet brane bound states \cite{pol} 
having no conventional space time interpretation. 

In this letter, we consider the effective action for 
strings and analyse the evolution of a four dimensional 
homogeneous isotropic universe with matter included. 
The matter is taken to be a perfect fluid with density 
$\rho (> 0)$, pressure $p$, and equation of state 
$p = \gamma \rho$, $- 1 \le \gamma \le 1$. We present 
a general analysis, applicable even in the absence of 
explicit solutions, from which the qualitative features 
of the evolution can be obtained. We find that 
the evolution, which is singular in general, 
becomes singularity free if during certain Phase of 
the evolution, when the scale factor increases and 
the effective string coupling becomes strong, 
the universe is dominated by solitonic $p$-branes, 
$p = 0$ and/or $- 1$, \footnote{$(- 1)$-branes are 
instantons but are also referred to as solitons here 
for the sake of brevity.} or by `matter' for which 
$\gamma \le - \frac{1}{\sqrt{3}}$. Further 
evolution is described in detail. 

It appears that either, or both, of these possibilities 
can be realised in string theory. The solitonic 
$p$-branes are heavy when the coupling is weak, but 
become light when the coupling is strong 
\cite{pol,witten}. Thus, they 
are likely to be produced copiously and dominate 
the universe during the relevent Phase. Using a result 
of Duff et al \cite{duff}, it then follows that 
the solitonic $p$-branes, $p = 0$ and/or $- 1$ do indeed 
avoid the singularity. This mechanism is reminiscent 
of the recently discovered field theory mechanism where 
heavy states - monopoles \cite{switt}, Ramond-Ramond 
black holes \cite{strom}, Dirichlet instantons 
\cite{ovafa} - become light and resolve the moduli 
space singularities. 

Alternately, when the scale factor increases, as during 
the relevent Phase, a gas of $p$-branes may have 
negative pressure. Indeed $\gamma = - \frac{1}{3}$ for 
a gas of strings \cite{bvafa}-\cite{ven1}. However, 
$\gamma$ is not known for arbitrary $p$, but it is 
plausible that $\gamma \le - \frac{1}{\sqrt{3}}$ for 
some $p$ (suggested to us by G. Veneziano). If true 
then the singularity can be avoided by such `matter' 
and the evolution becomes singularity free. 

The paper is organised as follows. We first present 
the action and the equations of motion with matter 
included. We rewrite the equations in a form suitable 
for our analysis. We then analyse the evolution in 
detail and obtain the conditions for the evolution 
to be completely singularity free. We then consider 
how a gas of solitonic $p$-branes may avoid 
the singularity and conclude with a few remarks. 

\vspace{2ex}

{\bf 2.}
Consider the string effective action for graviton 
and dilaton in the following form: 
\begin{equation}\label{schi}
S = - \frac{1}{2} \int d^4 x \sqrt{- g} \left( 
\chi R - \frac{\omega}{\chi} (\nabla \chi)^2\right) 
\end{equation}
where $g_{\mu \nu}$ is the string $\sigma$-model metric, 
$\chi \; (\ge 0)$ is the dilaton, and Newton's constant 
is set equal to $\frac{1}{8 \pi}$. The effective string 
coupling $g_s$ is given by $g_s = \frac{1}{\sqrt{\chi}}$ 
(the square roots are to be taken with a positive sign 
always). $\omega$ is a constant equal to $- 1$ for 
string theory but since the values of $\omega$ in 
the range $- \frac{3}{2} \le \omega < - 1$ will also 
be relevent in the following, we retain $\omega$ in 
(\ref{schi}) and assume only that $- \frac{3}{2} \le 
\omega \le - 1$. Action (\ref{schi}) can also be 
written in the canonical form 
\begin{equation}\label{sein}
S = - \frac{1}{2} \int d^4 x \sqrt{- \bar{g}} \left( 
\bar{R} - \frac{1}{2} (\bar{\nabla} \phi)^2\right) 
\end{equation}
where $\bar{g}_{\mu \nu} = \chi g_{\mu \nu}$ is 
the canonical metric and $\phi = \sqrt{2 \omega + 3} 
\; {\rm ln} \chi$. 

In this letter, we study the evolution of 
a flat \footnote{It is straightforward to incorporate 
spatial curvature also.} homogeneous isotropic universe 
with matter coupled minimally to $g_{\mu \nu}$ 
\cite{bvafa}-\cite{ven2},\cite{duff}. The matter is 
taken to be a perfect fluid with density $\rho$ $(> 0)$ 
and pressure $p$, related by the equation of state 
$p = \gamma \rho$ where $\gamma$ is a constant and 
$- 1 \le \gamma \le 1$. \footnote{The range of $\gamma$ 
includes the values corresponding to all known forms 
of matter such as, for example, vacuum energy density 
($\gamma = - 1$), dust ($\gamma = 0$), radiation 
($\gamma = \frac{1}{3}$), and massless scalar fields 
dust ($\gamma = 1$).} 
The fields depend on the time 
coordinate $t$ only and the line element is given by 
$d s^2 = - d t^2 + e^{2 A(t)} \left( d r^2 + r^2 
d \theta^2 + r^2 \sin^2 \theta d \varphi^2 \right)$ 
where $e^A$ is the scale factor. Defining $\Omega \equiv 
2 \omega + 3$ and using $p = \gamma \rho$, the equations 
of motion become 
\begin{eqnarray}
\dot{A} & = & - \frac{\dot{\chi}}{2 \chi} 
+ \sqrt{\frac{\rho}{6 \chi} 
+ \frac{\Omega \dot{\chi}^2}{12 \chi^2}} \label{ad} \\ 
\ddot{\chi} & = & - 3 \dot{A} \dot{\chi} 
+ \frac{(1 - 3 \gamma) \rho}{2 \Omega} \label{chidd} \\ 
\rho & = & \rho_0 e^{- 3 (1 + \gamma) A} \label{rho} 
\end{eqnarray} 
where upper dots denote $t$-derivatives and $\rho_0$ is 
a positive constant. The second term in (\ref{ad}) is 
taken to be positive so that one obtains an expanding 
universe ($\dot{A} > 0$) of the standard cosmology 
when $\dot{\chi} = 0$ identically. 

The analysis of the evolution is straightforward if 
equations (\ref{ad})-(\ref{rho}) can be solved explicitly 
for all times with $\rho$ and $\gamma$ arbitrary. Since 
this is not possible, \footnote{However, asymptotic power 
law solutions can be obtained \cite{ke}.} we adapt 
a different method \cite{k1,k2}. Note that equation 
(\ref{chidd}) can be integrated once to obtain 
\begin{equation}\label{chid}
\dot{\chi} (t) = e^{- 3 A} \; (\sigma (t) + c) 
\; , \; \; \; \; \sigma (t) \equiv \frac{(1 - 3 \gamma) 
\rho_0}{2 \Omega} \int^t_{t_i} dt e^{- 3 \gamma A} 
\end{equation}
and $c = \dot{\chi} e^{3 A} (t_i)$ is a constant. Then, 
dividing equation (\ref{ad}) by $\frac{\dot{\chi}}{\chi}$ 
and substituting (\ref{chid}) for $\dot{\chi}$, 
we obtain an equation relating $A$ and $\chi$: 
\begin{equation}\label{achi}
2 \chi \; \frac{d A}{d \chi} = - 1 + {\rm sign} 
(\dot{\chi}) \; \sqrt{K} \; , \; \; \; \; K \equiv 
\frac{\Omega}{3} \; \left( 1 + \frac{2 \rho_0 \chi 
e^{3 (1 - \gamma) A}}{(\sigma (t) + c)^2} \right) \; .  
\end{equation}
Also note that all the curvature invariants are finite 
and, hence, the singularities are absent if $e^{- A}$ 
and $\frac{d^n A}{d t^n}, \; n \ge 1$ or, equivalently 
as follows from the repeated use of (\ref{ad}) - 
(\ref{rho}), if 
\begin{equation}\label{qty}
e^{- A}, \; \frac{\rho}{\chi}, \; \; {\rm and} \; \; 
\frac{\dot{\chi}}{\chi} 
\end{equation}
all remain finite \cite{k1,k2}. Using equations 
(\ref{rho})-(\ref{achi}) and this sufficiency 
condition, the evolution can be analysed completely 
and the presence/absence of singularities determined 
even in the absence of explicit solutions. 

\vspace{2ex}

{\bf 3.}
We start at an initial time $t_{initial} \equiv 0$, 
corresponding to a temperature say $\stackrel{>}{_\sim} 
10^{16}$ GeV. This is so that various model dependent 
phenomena such as Grand Unification symmetry breaking, 
inflation, etc., which are in any case not relevent to 
the issue of singularity, may all occur 
for $t > 0$ only. For initial conditions we choose 
$\dot{A} (0) > 0$, $\chi (0) > 0$, and 
$\dot{\chi} (0) > 0$ \footnote{If $\dot{\chi} (0) < 0$ 
one merely starts in Phase 4 (see below).} 
corresponding to the fact that the universe is expanding 
and the string coupling is decreasing for $t \ge 0$. 
Their values, assumed to be non infinitesimal, are not 
required in the following. Also, $\chi (0)$ is assumed 
to be finite. 

For $t > 0$, the scale factor $e^A$ and the dilaton 
$\chi$ increase. In this era, various model dependent 
phenomena such as Grand Unification symmetry breaking, 
inflation, etc. may occur. However, 
they do not lead to singularity in any of the known 
models. Moreover, $e^A$ and $\chi$ both continue to 
increase during and after these phenomena. Eventually 
$\chi \to \infty$ and $e^A \to \infty$ and, as $t \to 
\infty$, the asymptotic solution that must describe 
the present era is given by 
\begin{equation}\label{present}
e^A = e^{A_0} t^n \; , \; \; \; \; 
\chi = \chi_0 t^m 
\end{equation}
where $A_0$ and $\chi_0$ are constants and $(n, m) 
= (\frac{1}{2}, - \frac{1}{2})$ in the radiation 
dominated era ($\gamma = \frac{1}{3}$) and 
$= (\frac{2 \omega + 2}{3 \omega + 4}, 
\frac{2}{3 \omega + 4})$ in the dust dominated era 
($\gamma = 0$) \cite{ke}. 

But this cannot be the entire story. Note that 
in string theory $\omega = - 1$ but experimental 
observations require that $\omega > 500$. This 
contradiction is avoided if string theory generates 
a potential for $\chi$ and thereby, or otherwise, 
freezes its dynamics. Such a mechanism, however, is 
not expected to lead to any singularity. Therefore, 
in this letter, we assume the existence of such 
a mechanism and will not pursue it further. 

\vspace{2ex}

{\bf 3 a.} 
Consider $t < 0$. Let $t' \equiv - t$ so that $t' > 0$ in 
this era. Then, in terms of $t'$, the initial conditions 
are $\dot{A} (0) < 0$, $\chi (0) > 0$, and $\dot{\chi} 
(0) < 0$ where upper dots now denote $t'$-derivatives. 
Equation (\ref{rho}) remains unchanged whereas equations 
(\ref{ad}), (\ref{chid}), and (\ref{achi}) become 
\begin{eqnarray}
\dot{A} & = & - \frac{\dot{\chi}}{2 \chi} 
- \sqrt{\frac{\rho}{6 \chi} 
+ \frac{\Omega \dot{\chi}^2} {12 \chi^2}} \label{ad'} \\ 
\dot{\chi} (t') & = & e^{- 3 A} \; 
(\sigma (t') + c) \label{chid'} \\ 
2 \chi \; \frac{d A}{d \chi} & = & - 1 - {\rm sign} 
(\dot{\chi}) \; \sqrt{K} \label{achi'} 
\end{eqnarray}
where the constant $c = \dot{\chi} e^{3 A} (0)$ is 
negative and 
\begin{eqnarray} 
\sigma (t') & = & \frac{(1 - 3 \gamma) \rho_0}{2 \Omega} 
\int^{t'}_0 dt' e^{- 3 \gamma A} \label{sigma'} \\ 
K & = & \frac{\Omega}{3} \; \left( 1 + \frac{2 \rho_0 
\chi e^{3 (1 - \gamma) A}}{(\sigma (t') + c)^2} 
\right) \; . \label{k} 
\end{eqnarray} 
It follows from the initial conditions that $\chi 
\frac{d A}{d \chi} (0) = \chi \frac{\dot{A}}{\dot{\chi}} 
(0) > 0$ and, hence from (\ref{achi'}), that $K (0) > 1$. 

Consider the evolution for $t' > 0$. Clearly, for 
$t' > 0$, the universe is dominated by radiation 
($\gamma = \frac{1}{3}$) or, when $e^A$ is 
sufficiently small, by massless scalar fields  
($\gamma = 1$). Therefore, $(1 - 3 \gamma) \le 0$. As 
$t'$ increases $e^A$ decreases and, hence, the integral 
in (\ref{sigma'}) increases. This implies, since 
$(1 - 3 \gamma) \le 0$, that $\sigma (t')$ decreases or 
remains constant. Therefore, $(\sigma (t') + c) \le c 
< 0$. Consequently, as $t'$ increases, $\dot{\chi} < 0$ 
and, hence, $\chi$ decreases. 

Since $e^A$ and $\chi$ decrease, and 
$(\sigma (t') + c)^2$ increases or remains constant, it 
follows that $K$ decreases monotonically. Its lowest 
value is $\frac{\Omega}{3} \le \frac{1}{3}$, achievable 
when $\chi$ vanishes, see (\ref{k}). Since $K (0) > 1$, 
it then follows that there exists a time, say $t' = t'_m 
> 0$ where $K (t'_m) = 1$ with $\chi (t'_m) > 0$. 
Therefore, $\frac{d A}{d \chi} (t'_m) = 0$ implying that 
$\dot{A} (t'_m) = 0$. This is a critical point of $e^A$ 
and is a minimum. Also, equation (\ref{achi'}) gives 
\begin{equation}\label{afinite}
A (t'_m) - A (0) = \int^{\chi (t'_m)}_{\chi (0)} 
\frac{d \chi}{2 \chi} \; (- 1 + \sqrt{K}) 
= {\rm finite} \; , 
\end{equation}
where the last equality follows because both 
the integrand and the interval of integration are finite. 
This implies that $A (t'_m)$ is finite and, therefore, 
that $e^{A (t'_m)}$ is finite and non vanishing. 

It can be seen that the quantities in (\ref{qty}) 
are all finite implying that the curvature invariants 
are all finite. Thus, there is no singularity for 
$0 \le t' \le t'_m$. 

\vspace{2ex}

{\bf 3 b.}
Let $t'_1 \equiv t'_m + \delta$ where $\delta$ is 
a positive infinitesimal constant. 
Then, by continuity, we have $\dot{A} (t'_1) > 0$, 
$\dot{\chi} (t'_1) < 0$, $(\sigma (t'_1) + c) < 0$, 
and $K (t'_1) < 1$. Thus, for $t' > t'_1$, $\chi$ 
decreases and $e^A$ increases. However, nothing 
can be said about $(\sigma (t') + c)$ or $K(t')$. 

To proceed further, assume that $K (t') < 1$ for all 
$t' > t'_1$. This necessarily requires that $\chi e^{3 
(1 - \gamma) A}$ remain finite and, since $(\sigma 
(t'_1) + c) < 0$, that $(\sigma (t') + c)$ remain 
negative and non infinitesimal. Therefore, $\dot{\chi} 
< 0$ and $\chi$ decreases continuosly whereas $\dot{A} 
> 0$ and $e^A$ increases continuosly. Then, in the limit 
$\chi \to 0$, equations (\ref{ad'})-(\ref{achi'}) can be 
solved explicitly. The solution is \cite{ke,ohan,k2} 
\begin{equation}\label{chi0}
{\rm for} \; \Omega \ne \frac{1}{3}: \; \; \; \; 
e^A = e^{A_0} \left( t'_s - {\rm sign} (m) t' \right)^n 
\; , \; \; \; \; \chi = \chi_0 
\left( t'_s - {\rm sign} (m) t' \right)^m \; , 
\end{equation}
where $A_0$, $\chi_0 > 0$, and $t'_s > t'_1$ are 
constants, and 
\begin{equation}\label{mn} 
n = \frac{3 - \sqrt{3 \Omega}}
{3 (1 - \sqrt{3 \Omega})} \; , \; \; \; \; 
m = \frac{- 2}{1 - \sqrt{3 \Omega}} \; ; 
\end{equation}
\begin{equation}\label{chi0e}
{\rm for} \; \Omega = \frac{1}{3}: \; \; \; \; 
e^A = e^{A_0} e^{k t'} \; , \; \; \; \; 
\chi = \chi_0 e^{- 3 k t'} \; , 
\end{equation}
where $k > 0$ is a constant. 

If $\Omega > \frac{1}{3}$ then $m > 0$ and $n < 0$. 
Thus, as $\chi \to 0$, $t' \to t'_s$ and 
$e^A \to \infty$, and it can be seen that 
the curvature scalar diverges. Therefore, there 
is a singularity at a finite time $t'_s$.  

If $\Omega < \frac{1}{3}$ then $m < 0$ and $n > 0$. 
Thus, as $\chi \to 0$, $t' \to \infty$  and $e^A \to 
\infty$. It can then be seen that the quantities in 
(\ref{qty}) are all finite for $t'_1 \le t' \le \infty$, 
implying that all the curvature invariants are finite 
and, hence, there is no singularity. This is true for 
$\Omega = \frac{1}{3}$ also. 

For strings, $\Omega = 1$ and therefore the solution is 
given by (\ref{chi0}) with $n = - \frac{1}{\sqrt{3}}$ 
and $m = \sqrt{3} + 1$. Thus there is a singularity 
at a finite time $t'_s$. \footnote{This singularity is 
similar to the one encountered in the superinflationary 
branch of the pre big bang cosmology \cite{ven1,ven2} 
where it leads to the graceful exit problem which, 
perhaps, may also be solved by the resolution to be 
given below.} 

Is this singularity unavoidable? Equivalently, is 
$K (t') < 1$ for all $t' > t'_1$? Consider the case 
where $\Omega = 1$ and, as $e^A \to \infty$, 
the universe is dominated by `matter' for which 
$\gamma \le - \frac{1}{\sqrt{3}}$. \footnote{More 
generally, by `matter' for 
which $\gamma \le \frac{1 - \sqrt{3 \Omega}} 
{3 - \sqrt{3 \Omega}}$ - thus, $\gamma \le 0$ for 
$\Omega = \frac{1}{3}$ and $\gamma \le \frac{1}{3}$ for 
$\Omega = 0$. When such `matter' is present the evolution 
with $\Omega < 1$ also proceeds as described below, and 
not as given by equations (\ref{chi0})-(\ref{chi0e}).} 
Then $(1 - 3 \gamma) > 0$ and, as $t' \to t'_s$, 
it can be seen easily that $\sigma (t') \to \infty$ 
(logarithmically if $\gamma = - \frac{1}{\sqrt{3}}$). 
Also, $\chi e^{3 (1 - \gamma) A} (t') \to \infty$. This 
implies that $(\sigma (t') + c)$ which is negative 
initially at $t'_1$ must cross zero before $t'_s$ and, 
hence, $K (t')$ must diverge. Since $K (t'_1) < 1$, it 
thus follows that there exists a time, say $t' = t'_M 
\; (t'_1 < t'_M < t'_s)$, where $K (t'_M) = 1$ and, 
hence, $\chi \frac{d A}{d \chi} (t'_M) = 0$. It also 
follows necessarily that $e^{A (t'_M)} < \infty$ (see 
(\ref{chi0})), $\chi (t'_M) > 0$ (by reversing 
the argument which led to (\ref{afinite})), 
$(\sigma (t'_M) + c) < 0$ (otherwise $K$ diverges), 
and $\dot{\chi} (t'_M) \ne 0$ (see (\ref{chid'})). 
Therefore, it now follows that $\dot{A} (t'_M) = 0$. 
This is a critical point of $e^A$ and is a maximum. 

It can be seen that the quantities in (\ref{qty}) 
are all finite for $t'_1 \le t' \le t'_M$, implying 
that all the curvature invariants are finite. Thus, 
there is no singularity for $t'_1 \le t' \le t'_M$. 

\vspace{2ex}

{\bf 3 c.}
Let $t'_2 \equiv t'_M + \delta$. 
Then, by continuity, we have $\dot{A} (t'_2) < 0$, 
$\dot{\chi} (t'_2) < 0$, $(\sigma (t'_2) + c) < 0$, 
and $K (t'_2) > 1$. If $(\sigma (t'_2) + c)$ remains 
negative and does not vanish for $t' > t'_2$ then 
the initial conditions at $t'_2$ are same as those 
at $t' = 0$ and, hence, the evolution proceeds as 
in {\bf 3 a}. Thus, the evolution becomes cyclical 
but remains singularity free. The dilaton $\chi$ 
decreases continuosly but remains finite and non 
vanishing for $t' < \infty$. 

If $(\sigma (t'_2) + c)$ vanishes at time, 
\footnote{This is likely to be the case since, 
as clear from the evolution for $t' \le t'_M$ 
in {\bf 3 b}, $(\sigma (t') + c)$ is increasing 
rapidly at $t'_M$.} say $t' = t'_n > t'_2$ then 
it implies that $\dot{\chi} (t'_n)  = 0$. This is 
a critical point of $\chi$ and is a minimum. 

It can be seen that the quantities in (\ref{qty}) 
are all finite for $t'_2 \le t' \le t'_n$, implying 
that all the curvature invariants are finite. Thus, 
there is no singularity for $t'_2 \le t' \le t'_n$. 

\vspace{2ex}

{\bf 3 d.}
Let $t'_3 \equiv t'_n + \delta$. 
Then, by continuity, we have $\dot{A} (t'_3) < 0$, 
$\dot{\chi} (t'_3) > 0$, and $(\sigma (t'_3) + c) > 0$. 
As $t'$ increases, $\chi$ increases and $e^A$ decreases 
and, in the absence of any other effects, $e^A$ would 
vanish at time, say $t' = t'_z > t'_3$. 

However, it follows from (\ref{rho}) that when $e^A$ 
is sufficiently small the universe is dominated by 
massless scalar fields since $\gamma = 1$ for them. 
Then $(1 - 3 \gamma) < 0$ and $\sigma (t')$, and 
hence $(\sigma (t') + c)$, begin 
to decrease. As $\chi \to \infty$, it follows that 
$\frac{\rho}{\chi}$-term dominates 
$\frac{\dot{\chi}}{\chi}$-terms in equation (\ref{ad'}). 
This implies that $e^A$ decreases faster than 
$(t'_z - t')^{\frac{1}{3}}$ as $t' \to t'_z$. 
Hence, $\sigma (t') \to - \infty$ faster than ${\rm ln} 
(t'_z - t')$ implying that $(\sigma (t') + c)$ 
which is positive initially at $t'_3$ must vanish at 
time, say $t' = t'_N \; (t'_3 < t_N < t'_z)$. 
Consequently $\dot{\chi} (t'_N) = 0$. This is 
a critical point of $\chi$ and is a maximum. 

It can be seen that the quantities in (\ref{qty}) 
are all finite for $t'_3 \le t' \le t'_N$, implying 
that all the curvature invariants are finite. Thus, 
there is no singularity for $t'_3 \le t' \le t'_N$. 

\vspace{2ex}

{\bf 3 e.}
Let $t'_4 \equiv t'_N + \delta$. 
Then, by continuity, we have $\dot{A} (t'_4) < 0$ and 
$\dot{\chi} (t'_4) < 0$. Also, $(\sigma (t'_4) + c) < 0$, 
and $K (t'_4) > 1$. Therefore, the initial conditions 
at $t'_4$ are same as those at $t' = 0$ and, hence, 
the evolution proceeds as described in {\bf 3 a}. Thus, 
the evolution becomes cyclical but remains singularity 
free. Also, during the course of the evolution, $\chi$ 
remains finite and non vanishing. 

We now summarise the evolution (we use the original 
time variable $t \equiv - t'$ and upper dots now 
denote $t$-derivatives).  \\
{\bf Phase 0 ($t \ge 0$):} 
In this phase, various model dependent phenomena such 
as Grand Unification symmetry breaking, inflation, 
generation of dilaton potential, etc. may occur but they 
do not lead to singularities. For $t \le 0$ we have 

\noindent
{\bf Phase 1 ($- t_1 \le t \le 0$):} 
$\dot{A} (0) > 0$, $\dot{\chi} (0) > 0$. As $t$ 
decreases, both $e^A$ and $\chi$ decrease. Then 
radiation ($\gamma = \frac{1}{3}$) or massless scalar 
fields ($\gamma = 1$) dominate the universe as $e^A$ 
becomes small. Under these conditions, $e^A$ always 
reaches a non zero minimum at time say $- t_1 + \delta$ 
where $\delta$ is a positive infinitesimal constant. 
For $t \le - t_1$ it leads to 

\noindent
{\bf Phase 2 ($- t_2 \le t \le - t_1$):} 
$\dot{A} (- t_1) < 0$, $\dot{\chi} (- t_1) > 0$. As $t$ 
decreases, $e^A$ increases and $\to \infty$ and $\chi$ 
decreases and $\to 0$. If $\Omega \le \frac{1}{3}$ then 
$e^A \to \infty$ and $\chi \to 0$ as $t \to - \infty$ 
and there is no singularity. If $\Omega = 1$ as in 
(\ref{schi}) then $e^A \to \infty$ and $\chi \to 0$ in 
a finite time and there is a singularity. However, in 
both cases, if there exists `matter' for which $\gamma 
\le \frac{1 - \sqrt{3 \Omega}}{3 - \sqrt{3 \Omega}}$ 
then it dominates the universe as $e^A$ becomes large. 
Under these conditions, $e^A$ always reaches a finite 
maximum at time say $- t_2 + \delta$, thereby avoiding 
the singularity in the case of $\Omega = 1$ (and, more 
generally, in the case of $\Omega > \frac{1}{3}$ also). 
For $t \le - t_2$ it leads to 

\noindent
{\bf Phase 3 ($- t_3 \le t \le - t_2$):} 
$\dot{A} (- t_2) > 0$, $\dot{\chi} (- t_2) > 0$. 
As $t$ decreases, $e^A$ and $\chi$ continue to decrease. 
$\chi$ may or may not reach a non zero minimum. In 
the later case, Phase 3 is identical to Phase 1. In 
the former case, let $\chi$ reach a minimum at time 
say $- t_3 + \delta$. For $t \le - t_3$ it leads to 

\noindent
{\bf Phase 4 ($- t_4 \le t \le - t_3$):} 
$\dot{A} (- t_3) > 0$, $\dot{\chi} (- t_3) < 0$. 
As $t$ decreases, $\chi$ increases and $e^A$ decreases. 
Then massless scalar fields ($\gamma = 1$) dominate 
the universe as $e^A$ becomes small. Under these 
conditions, $\chi$ always reaches a finite maximum at 
time say $- t_4 + \delta$. For $t \le - t_4$ it leads to 

\noindent
{\bf Phase 5} the initial conditions at the beginning 
of which are the same as those in Phase 1. Hence, 
the evolution also proceeds as described in Phase 1. 

\vspace{2ex}

{\bf 4.} 
The evolution is thus completely singularity free under 
the conditions given in Phase 2 during which $e^A$ 
increases, $\chi$ decreases and, hence, the effective 
string coupling $g_s = \frac{1}{\sqrt{\chi}}$ increases. 
During this Phase, if $\Omega$ becomes $\le \frac{1}{3}$ 
and if no `matter' exists for which $\gamma \le 
\frac{1 - \sqrt{3 \Omega}}{3 - \sqrt{3 \Omega}}$ then 
$e^A \to \infty$ and $\chi \to 0$ as $t \to - \infty$, 
given by (\ref{chi0})-(\ref{chi0e}); if such `matter' 
exists then, for any $\Omega \le 1$, both $e^A$ and 
$\chi$ evolve cyclically and remain finite 
and non vanishing as $t \to - \infty$. 

It appears that either or both of these possibilities, 
namely $\Omega$ becomes $\le \frac{1}{3}$ and/or 
`matter' exists for which $\gamma \le \frac{1 - 
\sqrt{3 \Omega}}{3 - \sqrt{3 \Omega}}$, can be realised 
in string theory by a gas of $p$-branes - the string 
solitons. To begin with, it is highly plausible 
that a gas of $p$-branes dominate the universe in 
Phase 2. Note that in string units, $p$-branes have 
masses $\sim \frac{1}{g_s}$ or $\frac{1}{g^2_s}$ 
\cite{pol,witten}. Hence, as $g_s$ becomes large, 
$p$-branes become light and, hence, are likely to be 
produced copiously thus dominating the universe 
during Phase 2. 

Duff {\em et al} have derived the metric 
$\tilde{g}_{\mu \nu}$ to which the (solitonic) 
$p$-branes couple minimally \cite{duff}. In terms 
of $\tilde{g}_{\mu \nu}$ and the $p$-brane dilaton 
$\tilde{\chi}$, related to $g_{\mu \nu}$ and $\chi$ by 
\begin{equation}\label{pmetric}
g_{\mu \nu} = \tilde{\chi}^{1 + \frac{p + 1}{\sqrt{p^2 
+ 3}}} \tilde{g}_{\mu \nu} \; , \; \; \; \; \chi = 
\tilde{\chi}^{- \frac{p + 1}{\sqrt{p^2 + 3}}} \; ,  
\end{equation}
the graviton-dilaton action can be written as in 
(\ref{schi}) but now with a $\omega$ given by 
\footnote{In a sense the change of $\omega$ is analogous 
to, but more involved than, changing $\gamma$ from $0$ 
to $\frac{1}{3}$ in standard cosmology when one goes 
from dust dominated to radiation dominated universe. As 
in the standard cosmology, here too the intermediate 
dynamics is easy to study but it is expedient to 
simply switch from one value of $\omega$ to other.} 
\begin{equation}\label{pomega}
\tilde{\Omega} \equiv 2 \tilde{\omega} + 3 
= \frac{(p + 1)^2}{p^2 + 3} \; . 
\end{equation} 

Note that for $p = 0$ and $- 1$, $\tilde{\Omega} 
= \frac{1}{3}$ and $0$ respectively. The evolution can 
now be analysed as in {\bf 3 b} \footnote{For $p = -1$, 
we first set $p = - 1 + \delta$ and, after performing 
the analysis, take the limit $\delta \to 0$.} and it 
follows that there is no singularity for these values of 
$\Omega$. If there also exists `matter' for which 
$\gamma \le \frac{1 - \sqrt{3 \tilde{\Omega}}} 
{3 - \sqrt{3 \tilde{\Omega}}}$ then further evolution 
proceeds as described in {\bf 3 b - 3 e}: the scale 
factor and the dilaton evolve cyclically and both 
remain finite and non vanishing. 

The heavy solitonic $p$-branes becoming light and 
produced copiously, thus resolving the singularity is 
reminiscent of the recently discovered field theory 
mechanism \cite{switt}-\cite{ovafa} where heavy states 
- monopoles in \cite{switt}, Ramond-Ramond black holes 
in \cite{strom}, Dirichlet instantons in \cite{ovafa} 
- become light and resolve the moduli 
space singularities. Also, the strong coupling regime 
of the original theory ($\chi$ small) corresponds to 
the weak coupling regime of the solitons 
\footnote{The effective $p$-brane coupling can be seen, 
by using the action (\ref{schi}) and by an analogy with 
the effective string coupling, to be given by 
$\tilde{g}_p = \frac{1}{\sqrt{\tilde{\chi}}}$.} 
($\tilde{\chi}$ large), see (\ref{pmetric}). 

Alternately, when $e^A$ is increasing, a gas of 
$p$-branes may have negative pressure, and hence 
negative $\gamma$. In fact, $\gamma$ is shown to be 
$= - \frac{1}{3}$ for strings ($p = 1$)  
\cite{bvafa}-\cite{ven1}. Tseytlin considers 
the case $- \frac{1}{\sqrt{3}} < \gamma < 
\frac{1}{\sqrt{3}}$ also \cite{tseyt}, although 
the nature of the corresponding `matter' is not 
clear. (See \cite{horowitz} also for another (higher 
dimensional) context where a sufficient negative 
pressure may arise due to Dirichlet branes.) In 
the present case, the value of $\gamma$ for arbitrary 
$p$ is not known, but it is plausible that $\gamma \le 
- \frac{1}{\sqrt{3}}$ for some $p$ (suggested to us by 
G. Veneziano). If true then again the singularity in 
Phase 2 can be avoided and the evolution becomes 
singularity free. 

Thus, the resulting scenario for the resolution of 
cosmological singularity is that as the string coupling 
becomes strong in Phase 2, solitonic $p$-branes are 
produced copiously and dominate the universe. They 
either change $\omega$ in (\ref{schi}) from $- 1$ to 
$\tilde{\omega}$ given in (\ref{pomega}) or produce 
a negative pressure corresponding to $\gamma \le 
- \frac{1}{\sqrt{3}}$ or both. The singularity 
in Phase 2 is then avoided and further evolution 
is singularity free. 

\vspace{2ex}

{\bf 5.} 
A few remarks are now in order. First, we have considered 
various metrics - the string $\sigma$-model metric 
$g_{\mu \nu}$, the canonical metric $\bar{g}_{\mu \nu}$, 
and the $p$-brane metric $\tilde{g}_{\mu \nu}$ which are 
related to each other through dilaton dependent conformal 
factors (see (\ref{sein}) and (\ref{pmetric})). Therefore 
when one metric is singularity free then the other two 
are also singularity free if the dilaton remains finite 
and non vanishing, which is ensured if `matter' with 
appropriate $\gamma$ exists (see Phase 2). If $p$-branes, 
$p = 0$ and/or $- 1$, dominate the universe during 
Phase 2 but `matter' with appropriate $\gamma$ does 
not exist then $p$-brane metric $\tilde{g}_{\mu \nu}$ 
evolves as in (\ref{chi0}) or (\ref{chi0e}) and is 
singularity free. It can be shown, following the methods 
of \cite{k1}, that any metric $\hat{g}_{\mu \nu} \equiv 
\tilde{\chi}^{\alpha} \tilde{g}_{\mu \nu}$ is 
singularity free and the corresponding time can be 
continued indefinitely into the past and the future if 
and only if $\alpha \le 1 - \sqrt{3 \tilde{\Omega}}$. It 
therefore follows from (\ref{pomega}) and (\ref{pmetric}) 
that when $0$-branes dominate the universe 
$\tilde{g}_{\mu \nu}$ is singularity free but 
$g_{\mu \nu}$ and $\bar{g}_{\mu \nu}$ are singular 
whereas when $(- 1)$-branes dominate the universe 
all of these metrics are singularity free. 

Second, action (\ref{schi}) is unlikely to be valid 
in the strong coupling regime. This is just as well 
since otherwise the evolution is singular and 
the singularities cannot be avoided. However, 
an effective action valid at strong coupling is not 
known. In a sense the above scenario, which ensures 
singularity free evolution of the universe, can 
be viewed as a conjecture towards obtaining such 
an action. Or, given the analogy between the above 
scenario and the resolution of moduli space 
singularities, the methods of \cite{switt}-\cite{ovafa} 
can perhaps be applied to understand the details 
of strong coupling effects near the cosmological 
singularities also.  

Third, action (\ref{schi}) may receive higher derivative 
corrections, which must be included. However, when 
the curvature invariants all remain finite they may not 
be crucial to the cosmological evolution. Also, 
in string theory such corrections appear to ameliorate 
the singularity problem \cite{art}-\cite{brand}, so it 
is possible that they improve the singularity aspects 
in the present case also. However, further study 
is required to understand the effects, crucial or 
otherwise, of higher derivative terms on cosmological 
evolution. Important though such a study is, we defer it 
to future as it is beyond the scope of the present work. 

{\bf Acknowledgement:} We thank D. P. Jatkar and 
G. Veneziano for discussions. 



\end{document}